\documentstyle[12pt]{article}
\def\be{\begin{equation}}
\def\ee{\end{equation}}
\def\bea{\begin{eqnarray}}
\def\eea{\end{eqnarray}}

\begin{document}

\begin{center}
{\Large{\bf Conformal Field Theory with Two Kinds of Bosonic Fields 
and Two Linear Dilatons}}

\vskip .5cm
{\large Davoud Kamani}
\vskip .1cm
{\it Faculty of
 Physics, Amirkabir University of Technology (Tehran Polytechnic)\\
 P.O.Box: 15875-4413, Tehran, Iran}\\
{\it e-mail: kamani@aut.ac.ir}\\
\end{center}

\begin{abstract}

We consider a two-dimensional conformal field theory
which contains two kinds of the bosonic degrees of freedom.
Two linear dilaton fields enable us to study a more general case.
Various properties of the model such as OPEs, central charge,
conformal properties of the fields and associated algebras will be studied.

\end{abstract}
\vskip .5cm



\newpage
\section{Introduction}
 
Among the various conformal field theories (CFTs) 
the linear dilaton CFT has some interesting applications
in the string theory, $e.g.$ see \cite{1,2,3,4,5,6}. In this CFT, 
modification of the energy-momentum tensor was anticipated by a linear
dilaton field. This CFT significantly changes the 
behavior of the worldsheet theory. 
In addition, this CFT is a consistent way for reducing 
the spacetime dimension without compactification.
The linear dilaton CFT also appears as an ingredient in many
string backgrounds, critical and non-critical. 
According to its importance we proceed to develop it.

In this paper we consider an action which is generalization of the bosonic
part of the ${\cal{N}}=2$ superstring theory. In addition, we 
introduce two linear dilaton fields to build our model. Thus, we
study a conformally invariant field theory in two flat dimensions.
Anticipating to the string theory, we refer to these two dimensions as the
string worldsheet. Various OPEs of the model will be calculated. 
Due to the dilaton fields, some of these OPEs, and also conformal 
transformations of the worldsheet fields $X^\mu(\sigma , \tau)$
and $Y^\mu(\sigma , \tau)$ have deviations from the 
standard forms. Presence of some parameters in the central 
charge enables us to receive a desirable dimension for the spacetime.
The algebra of the oscillators reveals that the oscillators of $X$-fields
do not commute with that of $Y$-fields. 

This paper is organized as follows. In section 2, we shall introduce
the action of the model and the linear dilatonic 
energy-momentum tensor associated to it.
In section 3, various OPEs of the model will be studied. In section 4,
conformal transformations of the worldsheet fields will be obtained.
In section 5, various quantities will be expressed 
in terms of the oscillators,
and two algebras for the model will be obtained. Section 6 is devoted
for the conclusions. 
\section{The Model}

We begin with the action of the scalar fields 
$X^\mu (z, {\bar z})$ and $Y^\mu (z, {\bar z})$ in two dimensions
\bea
S= \frac{1}{2\pi \alpha'} 
\int d^2 z (\partial_z X^\mu \partial_{\bar z} X_\mu
+\beta \partial_z Y^\mu \partial_{\bar z} Y_\mu
+\lambda(\partial_z X^\mu \partial_{\bar z} Y_\mu
+\partial_{\bar z} X^\mu\partial_z Y_\mu)),
\eea
where $\mu \in \{0,1,...,D-1\}$ and 
$\beta$ and $\lambda$ are constants, $i.e.$ independent of $z$
and ${\bar z}$. For the spacetime we consider the flat Minkowski metric
$\eta_{\mu\nu}={\rm diag}(-1,1,...,1) $. 
The special case $\beta=1$ and $\lambda=0$ indicates the bosonic part
of the ${\cal{N}}=2$ super-conformal field theory.
Thus, we say the set $\{X^\mu \}$ describes the spacetime
coordinates. In other words, $X^\mu (\sigma , \tau)$ is regarded as the
embedding of the worldsheet in the spacetime. 
However, $Y^\mu (\sigma , \tau)$ enters essentially in the same way. 
So  the set $\{Y^\mu (\sigma , \tau)\}$ does not describe additional dimensions. 
The conformal invariance of this action, $i.e.$ symmetry under the conformal
transformations $z \rightarrow z'(z)$ and 
${\bar z} \rightarrow {\bar z}'({\bar z})$, leads to the
zero conformal weights for the fields $X^\mu$ and $Y^\mu$.

The equations of motion, extracted from the action (1), are
\bea
&~& \partial_z \partial_{\bar z}X^\mu 
+\lambda \partial_z \partial_{\bar z}Y^\mu=0,
\nonumber\\
&~& \lambda \partial_z \partial_{\bar z}X^\mu 
+\beta \partial_z \partial_{\bar z}Y^\mu=0.
\eea
We assume that the determinant of the coefficients of these equations
to be nonzero $i.e.$,
\bea
\det \left( \begin{array}{cc}
1 & \lambda \\
\lambda & \beta
\end{array} \right)
=\beta-\lambda^2 \neq 0.
\eea
Therefore, we obtain
\bea
&~& \partial_z \partial_{\bar z}X^\mu=0,
\nonumber\\
&~& \partial_z \partial_{\bar z}Y^\mu=0.
\eea
These imply $\partial_z X^\mu$ and $\partial_z Y^\mu$ are functions of $z$,
and $\partial_{\bar z} X^\mu$ and $\partial_{\bar z} Y^\mu$ are functions
of ${\bar z}$.

The corresponding energy-momentum tensor has the components
\bea
&~&T'_{zz} \equiv T'(z)=-\frac{1}{\alpha'}(:\partial_z X^\mu\partial_z X_\mu:
+\beta :\partial_z Y^\mu\partial_z Y_\mu:
+\lambda :\partial_z X^\mu\partial_z Y_\mu:
+\lambda :\partial_z Y^\mu\partial_z X_\mu:),
\nonumber\\
&~& T'_{{\bar z}{\bar z}} \equiv {\tilde T'}({\bar z})
=-\frac{1}{\alpha'}(: \partial_{\bar z} X^\mu\partial_{\bar z} X_\mu:
+\beta :\partial_{\bar z} Y^\mu \partial_{\bar z} Y_\mu:
+\lambda :\partial_{\bar z} X^\mu \partial_{\bar z} Y_\mu:
+\lambda :\partial_{\bar z} Y^\mu \partial_{\bar z} X_\mu:)\;,
\nonumber\\
&~& T'_{z{\bar z}}=T'_{{\bar z}z}=0,
\eea
where : : denotes normal ordering.

It is possible to construct a more general CFT with the same action (1),
but with different energy-momentum tensor
\bea
&~& T(z)= \Lambda_{ij} :\partial_z X_i^\mu \partial_z X_{j\mu}:
+ V_\mu^i \partial_z^2 X^\mu_i ,
\nonumber\\
&~& {\tilde T}({\bar z})
=\Lambda_{ij} :\partial_{\bar z} X_i^\mu 
\partial_{\bar z}X_{j\mu}:
+ V_\mu^i \partial_{\bar z}^2 X^\mu_i ,
\nonumber\\
&~& T_{z{\bar z}}=T_{{\bar z}z}=0,
\eea
where sum over $i$ and $j$ is assumed with $i,j \in \{1,2\}$. We define 
\bea
&~& X^\mu_1 =X^\mu\;\;,\;\;\;\;X^\mu_2 =Y^\mu,
\nonumber\\
&~& V^1_\mu =V_\mu\;\;\;\;,\;\;\;\;V^2_\mu =U_\mu,
\nonumber\\
&~& \Lambda=-\frac{1}{\alpha'}
\left( \begin{array}{cc}
1 & \lambda \\
\lambda & \beta
\end{array} \right).
\eea 
The vectors $V^\mu$ and $U^\mu$ are fixed in the spacetime. 
For each pair of these vectors we have a CFT. 
The extra terms in (6) are total derivatives. 
Thus, we shall see that they do not affect the 
status of $T(z)$ and ${\tilde T}({\bar z})$
as generators of conformal transformations.
The field $\Phi =V_\mu X^\mu$ in (6) is linear dilaton. 
In the same way $\Phi' = U_\mu Y^\mu$ is a linear field in the
Y-space. In fact, by introducing the worldsheet fields 
$\{Y^\mu (\sigma,\tau)\}$ and defining
the energy-momentum tensor (6), we have 
generalized the linear dilaton CFT. The case $\beta=1$ and $\lambda=0$
decomposes the model to two copies of the linear dilaton CFT.
\section{Operator Product Expansions (OPEs)}

\subsection{The OPEs $XX$, $XY$ and $YY$}

We use the path integral formalism to derive operator equations.
Since the path integral of a total derivative is zero, we have the equation
\bea
&~& 0=\int DX DY \frac{\delta}{\delta X_\mu (z ,{\bar z})}\bigg{[}
e^{-S}X^\nu (w ,{\bar w})...\bigg{]}
\nonumber\\
&~&=\int DX DY e^{-S}\bigg{[}\bigg{(}
\frac{1}{\pi \alpha'}\partial_z \partial_{\bar z}
[X^\mu(z ,{\bar z}) + \lambda Y^\mu(z ,{\bar z})] X^\nu (w ,{\bar w})
+\eta^{\mu\nu}\delta^{(2)}(z -w,
{\bar z}-{\bar w})\bigg{)}...\bigg{]} 
\nonumber\\
&~&=\langle \bigg{(}\frac{1}{\pi \alpha'}\partial_z \partial_{\bar z}
[X^\mu(z ,{\bar z}) 
+ \lambda Y^\mu(z ,{\bar z})]X^\nu (w ,{\bar w})+  
\eta^{\mu\nu}\delta^{(2)}(z -w,{\bar z}-{\bar w}) \bigg{)}
... \rangle.
\eea
The point $(w, {\bar w})$ might be coincident with $(z, {\bar z})$, but
the insertion ``$... $'' is arbitrary, which is away from
$(z ,{\bar z})$ and $(w ,{\bar w})$.
Arbitraryness of the insertion implies
\bea
\frac{1}{\pi \alpha'}\partial_z \partial_{\bar z}
[X^\mu(z ,{\bar z}) + \lambda Y^\mu(z ,{\bar z})]
X^\nu (w ,{\bar w})=-\eta^{\mu\nu}\delta^{(2)}(z -w,{\bar z}-{\bar w}),
\eea
as an operator equation. In the same way, the equation
\bea
\int DX DY \frac{\delta}{\delta X_\mu (z ,{\bar z})}\bigg{[}
e^{-S}Y^\nu (w ,{\bar w})...\bigg{]}=0,
\eea
gives the operator equation
\bea
\partial_z \partial_{\bar z}
[X^\mu(z ,{\bar z}) + \lambda Y^\mu(z ,{\bar z})]
Y^\nu (w ,{\bar w})=0.
\eea
In the equation (10) change $X_\mu$ to $Y_\mu$, we obtain
\bea
\frac{1}{\pi \alpha'}\partial_z \partial_{\bar z}
[\lambda X^\mu(z ,{\bar z}) + \beta Y^\mu(z ,{\bar z})]
Y^\nu (w ,{\bar w})=-\eta^{\mu\nu}\delta^{(2)}(z -w,{\bar z}-{\bar w}).
\eea
Similarly, in the first line of (8) replacing $X_\mu$ by $Y_\mu$ leads to 
\bea
\partial_z \partial_{\bar z}
[\lambda X^\mu(z ,{\bar z}) + \beta Y^\mu(z ,{\bar z})]
X^\nu (w ,{\bar w})=0.
\eea

The equations (9), (11), (12) and (13) give the following equations
\bea
&~& \partial_z \partial_{\bar z} X^\mu(z ,{\bar z}) X^\nu (w ,{\bar w})
=-\frac{\pi \beta \alpha'}{\beta -\lambda^2}\eta^{\mu\nu}
\delta^{(2)}(z -w,{\bar z}-{\bar w}),
\nonumber\\
&~& \partial_z \partial_{\bar z} Y^\mu(z ,{\bar z}) X^\nu (w ,{\bar w})
=\frac{\pi \lambda \alpha'}{\beta -\lambda^2}\eta^{\mu\nu}
\delta^{(2)}(z -w,{\bar z}-{\bar w}),
\nonumber\\
&~& \partial_z \partial_{\bar z} X^\mu(z ,{\bar z}) Y^\nu (w ,{\bar w})
=\frac{\pi \lambda \alpha'}{\beta -\lambda^2}\eta^{\mu\nu}
\delta^{(2)}(z -w,{\bar z}-{\bar w}),
\nonumber\\
&~& \partial_z \partial_{\bar z} Y^\mu(z ,{\bar z}) Y^\nu (w ,{\bar w})
=-\frac{\pi \alpha'}{\beta -\lambda^2}\eta^{\mu\nu}
\delta^{(2)}(z -w,{\bar z}-{\bar w}).
\eea
That is, the equations of motion (4) hold except at the coincident point
$(z, {\bar z})=(w, {\bar w})$.
Define the matrix $Q_{ij}$ as in the following
\bea
Q=\frac{\alpha'}{2(\beta-\lambda^2)}
\left( \begin{array}{cc}
\beta & -\lambda \\
-\lambda & 1
\end{array} \right).
\eea
Thus, the equations (14) can be written in the compact form
\bea
\partial_z \partial_{\bar z} X_i^\mu(z ,{\bar z}) X_j^\nu (w ,{\bar w})
=-2\pi Q_{ij}\eta^{\mu\nu}
\delta^{(2)}(z -w,{\bar z}-{\bar w}).
\eea
According to this, we have the normal ordered equation
\bea
:X_i^\mu (z ,{\bar z})X_j^\nu (w ,{\bar w}):
=X_i^\mu (z ,{\bar z})X_j^\nu (w ,{\bar w})
+Q_{ij} \eta^{\mu\nu} \ln |z - w|^2 ,
\eea 
where $\partial_z \partial_{\bar z}\ln |z -w|^2 
= 2\pi \delta^{(2)}(z -w,{\bar z}-{\bar w})$ has been used.
The equation (16) and (17) indicate the equation of motion
\bea
\partial_z \partial_{\bar z}:X_i^\mu (z ,{\bar z})X_j^\nu (w ,{\bar w}):=0.
\eea
\subsection{The $TT$ OPE}

Let ${\cal{F}}$ be any functional of $\{X^\mu\}$ and $\{Y^\mu\}$. 
Thus, the generalization of (17) is defined by
\bea
{\cal{F}}[X,Y]=\exp \bigg{(}-\frac{1}{2} Q_{ij}\int d^2 z_1d^2z_2
\ln |z_1-z_2|^2 \frac{\delta}{\delta X^\mu_i (z_1 , {\bar z}_1)}
\frac{\delta}{\delta X_{j\mu} (z_2 , {\bar z}_2)}
\bigg{)}:{\cal{F}}[X,Y]:\;.
\eea
The OPE for any pair of the operators ${\cal{F}}$ and ${\cal{G}}$
is given by
\bea
:{\cal{F}}::{\cal{G}}:=\exp \bigg{(}-Q_{ij}\int d^2z_1d^2z_2
\ln |z_1-z_2|^2 \frac{\delta}{\delta X^\mu_{iF} (z_1 , {\bar z}_1)}
\frac{\delta}{\delta X_{\mu G}^{j} (z_2 , {\bar z}_2)}
\bigg{)}:{\cal{F}}{\cal{G}}:\;,
\eea
where the functional derivatives act only on the fields 
${\cal{F}}$ or ${\cal{G}}$, respectively.
For ${\cal{F}}=X^\mu_i (z , {\bar z})$ and 
${\cal{G}}=X^\nu_j(w , {\bar w})$ this reduces to the equation (17),
as expected.

Using the equation (20), we obtain
\bea
&~& :\partial_z X^\mu_i(z)\partial_z X_{j\mu}(z):
:\partial_w X^\nu_k(w)\partial_w X_{l\nu}(w):=
\nonumber\\
&~& :\partial_z X^\mu_i(z)\partial_z X_{j\mu}(z)
\partial_w X^\nu_k(w)\partial_w X_{l\nu}(w):
+ \frac{D}{(z-w)^4}(Q_{ik}Q_{jl}+Q_{il}Q_{jk})
\nonumber\\
&~& -\frac{1}{(z-w)^2}\bigg{(}
Q_{ik}:\partial_z X^\mu_j(z)\partial_w X_{l\mu}(w):+
Q_{il}:\partial_z X^\mu_j(z)\partial_w X_{k\mu}(w):
\nonumber\\
&~& +Q_{jk}:\partial_z X^\mu_i(z)\partial_w X_{l\mu}(w):+
Q_{jl}:\partial_z X^\mu_i(z)\partial_w X_{k\mu}(w):\bigg{)}.
\eea
Now the Taylor expansion of $\partial_z X^\mu_i(z)$ around $z=w$ changes this
equation to
\bea
&~& :\partial_z X^\mu_i(z)\partial_z X_{j\mu}(z):
:\partial_w X^\nu_k(w)\partial_w X_{l\nu}(w):
\;\sim \frac{D}{(z-w)^4}(Q_{ik}Q_{jl}+Q_{il}Q_{jk})
\nonumber\\
&~& -\frac{1}{(z-w)^2}\bigg{(}
Q_{ik}:\partial_w X^\mu_j(w)\partial_w X_{l\mu}(w):+
Q_{il}:\partial_w X^\mu_j(w)\partial_w X_{k\mu}(w):
\nonumber\\
&~& +Q_{jk}:\partial_w X^\mu_i(w)\partial_w X_{l\mu}(w):+
Q_{jl}:\partial_w X^\mu_i(w)\partial_w X_{k\mu}(w):\bigg{)}
\nonumber\\
&~& -\frac{1}{z-w}\bigg{(}
Q_{ik}:\partial^2_w X^\mu_j(w)\partial_w X_{l\mu}(w):+
Q_{il}:\partial^2_w X^\mu_j(w)\partial_w X_{k\mu}(w):
\nonumber\\
&~& +Q_{jk}:\partial^2_w X^\mu_i(w)\partial_w X_{l\mu}(w):+
Q_{jl}:\partial^2_w X^\mu_i(w)\partial_w X_{k\mu}(w):\bigg{)},
\eea
where the non-singular terms have been omitted.
For calculating the $TT$ OPE we also need the following OPEs
\bea
&~& :\partial_z X^\nu_i(z)\partial_z X_{j\nu}(z):\partial^2_w X^\mu_k (w)
\sim 
\nonumber\\
&~& -\frac{2}{(z-w)^3} (Q_{ik}\partial_w X_j^\mu(w)
+Q_{jk}\partial_w X_i^\mu(w))
\nonumber\\
&~& -\frac{2}{(z-w)^2} (Q_{ik}\partial^2_w X_j^\mu(w)
+Q_{jk}\partial^2_w X_i^\mu(w))
\nonumber\\
&~& -\frac{1}{z-w} (Q_{ik}\partial^3_w X_j^\mu(w)
+Q_{jk}\partial^3_w X_i^\mu(w)),
\eea
\bea
&~& \partial^2_z X^\mu_k(z):\partial_w X^\nu_i(w) \partial_w X_{j\nu}(w):
\;\sim \frac{2}{(z-w)^3} (Q_{ki}\partial_w X_j^\mu(w)
+Q_{kj}\partial_w X_i^\mu(w)),
\eea
\bea
\partial^2_z X^\mu_i (z)\partial^2_w X^\nu_j (w) \sim 
\frac{6}{(z-w)^4}\eta^{\mu\nu} Q_{ij}.
\eea
Adding all these together we obtain the $TT$ OPE as in the following
\bea
&~& T(z)T(w) \sim \frac{c}{2(z-w)^4}-\frac{1}{(z-w)^2}\bigg{[}
2\Lambda_{ij}V^k_\mu \bigg{(} Q_{ik} \partial^2_w X^\mu_j(w)
+ Q_{jk}\partial^2_w X^\mu_i (w)\bigg{)}
\nonumber\\
&~& +\Lambda_{ij}\Lambda_{kl}\bigg{(} 
Q_{ik} \partial_w X_{j\mu}(w)\partial_w X^\mu_l(w)+
Q_{il} \partial_w X_{j\mu}(w)\partial_w X^\mu_k(w)
\nonumber\\
&~& +Q_{jk} \partial_w X_{i\mu}(w)\partial_w X^\mu_l(w)+
Q_{jl} \partial_w X_{i\mu}(w)\partial_w X^\mu_k(w) 
\bigg{)}\bigg{]}
\nonumber\\
&~& -\frac{1}{z-w} \bigg{[}\Lambda_{ij} \Lambda_{kl}
\bigg{(}Q_{ik} \partial^2_w X_{j\mu}(w)\partial_w X^\mu_l(w)+
Q_{il} \partial^2_w X_{j\mu}(w)\partial_w X^\mu_k(w)
\nonumber\\
&~& +Q_{jk} \partial^2_w X_{i\mu}(w)\partial_w X^\mu_l(w)+
Q_{jl} \partial^2_w X_{i\mu}(w)\partial_w X^\mu_k(w) \bigg{)}
\nonumber\\
&~& +\Lambda_{ij} V^k_\mu \bigg{(}Q_{ik}\partial^3_w X^\mu_j(w)
+Q_{jk}\partial^3_w X^\mu_i(w)\bigg{)}\bigg{]}.
\eea
The ${\tilde T}{\tilde T}$ OPE also has a similar form in terms of
${\bar z}$, ${\bar w}$ and ${\tilde c}$.

The central charges are given by
\bea
&~& c={\tilde c}=2D \Lambda_{ij}\Lambda_{kl}
(Q_{ik}Q_{jl}+Q_{il}Q_{jk})+12Q_{ij}V^i_\mu V^\mu_j ,
\nonumber\\
&~& =2D + \frac{6\alpha'}{\beta-\lambda^2}(\beta V_\mu V^\mu
-2\lambda V_\mu U^\mu + U_\mu U^\mu).
\eea
Vanishing conformal anomaly relates the parameters of the model.
That is, string actually can move in a wide
range of the dimensions. However, by adjusting the variables
$\beta$, $\lambda$, $V^\mu$ and $U^\mu$, we can obtain desirable
dimension for the spacetime. 
The case $V^\mu=U^\mu=0$ gives the central charges $c={\tilde c}=2D$,
$i.e.$, $D$ for $\{X^\mu\}$ and $D$ for $\{Y^\mu\}$.
If $\beta=1$ and $\lambda=0$, the action (1) reduces to two copies
of the free string action, and hence the energy-momentum tensor (6)
is modified to two copies of the energy-momentum tensor of the linear 
dilaton CFT. In this case the central charge also reduces to 
two copies of the central charge of the linear dilaton CFT
\bea
&~& c=c_X+c_Y,
\nonumber\\
&~& c_X=D+6\alpha' V_\mu V^\mu,
\nonumber\\
&~& c_Y=D+6\alpha' U_\mu U^\mu.
\eea  

Since there is $\Lambda Q=-\frac{1}{2}I_{2\times 2}$, 
the $TT$ OPE (26), and similarly the ${\tilde T}{\tilde T}$
OPE take the standard forms
\bea
T(z) T(w) \sim \frac{c}{2(z-w)^4}+\frac{2}{(z-w)^2}T(w)+\frac{1}{z-w}
\partial_w T(w),
\nonumber\\
{\tilde T}({\bar z}){\tilde T}({\bar w}) \sim 
\frac{{\tilde c}}{2({\bar z}-{\bar w})^4}+
\frac{2}{({\bar z}-{\bar w})^2}{\tilde T}({\bar w})+
\frac{1}{{\bar z}-{\bar w}}\partial_{\bar w}{\tilde T}({\bar w}).
\eea
According to the central charge terms, $T$ and ${\tilde T}$ are not
conformal tensors. Apart from these terms, (29) is the statement that
$T(z)$ and ${\tilde T}({\bar z})$ are conformal fields of the weights
(2,0) and (0,2), respectively.
\subsection{The OPEs $TX$, $TY$, ${\tilde T}X$ and ${\tilde T}Y$}

The OPE $TX_k^\mu$ is
\bea
T(z)X_k^\mu(w,{\bar w}) \sim \frac{1}{(z-w)^2}V^\mu_i Q_{ik}
-\frac{1}{z-w}\Lambda_{ij}[Q_{ik}\partial_w X^\mu_j(w)
+Q_{jk}\partial_w X^\mu_i(w)].
\eea
Thus, for $k=1$ and $k=2$ we obtain
\bea
&~& T(z)X^\mu(w,{\bar w}) \sim \frac{1}{(z-w)^2}
\frac{\alpha'}{2(\beta-\lambda^2)}
(\beta V^\mu -\lambda U^\mu) + \frac{1}{z-w}\partial_w X^\mu(w),
\nonumber\\
&~& T(z)Y^\mu(w,{\bar w}) \sim \frac{1}{(z-w)^2}
\frac{\alpha'}{2(\beta-\lambda^2)}
(-\lambda V^\mu +U^\mu) + \frac{1}{z-w}\partial_w Y^\mu(w).
\eea
In the same way we have
\bea
&~& {\tilde T}({\bar z})X^\mu(w,{\bar w}) 
\sim \frac{1}{({\bar z}-{\bar w})^2}\frac{\alpha'}{2(\beta-\lambda^2)}
(\beta V^\mu -\lambda U^\mu) + \frac{1}{{\bar z}-{\bar w}}
\partial_{\bar w} X^\mu({\bar w}),
\nonumber\\
&~& {\tilde T}({\bar z})Y^\mu(w,{\bar w}) 
\sim \frac{1}{({\bar z}-{\bar w})^2}\frac{\alpha'}{2(\beta-\lambda^2)}
(-\lambda V^\mu +U^\mu) + \frac{1}{{\bar z}-{\bar w}}
\partial_{\bar w} Y^\mu({\bar w}).
\eea

The $U$-terms and $V$-terms imply that $X^\mu$ and $Y^\mu$ are not
conformal tensor operators. Putting away these terms 
(the square singular terms) of the above OPEs leads to the conditions
\bea
&~& \beta V^\mu -\lambda U^\mu =0,
\nonumber\\
&~& -\lambda V^\mu +U^\mu=0.
\eea
Since we assumed $\beta -\lambda^2 \neq 0$, we obtain $V^\mu=U^\mu =0$.
Therefore, (31) and (32) reduce to the OPEs $T'X$, $T'Y$, 
${\tilde T'}X$ and ${\tilde T'}Y$. That is, with $T'$ and ${\tilde T'}$
the fields $X^\mu$ and $Y^\mu$ are conformal tensors, as expected.
However, we shall not consider the case (33).
\section{Conformal Transformations of $X^\mu$ and $Y^\mu$}

The infinitesimal conformal transformations 
$z \rightarrow z'=z+\epsilon g(z)$ and
${\bar z} \rightarrow {\bar z'}={\bar z}+\epsilon g(z)^*$ imply the currents
\bea
&~& j(z) = ig(z)T(z),
\nonumber\\
&~& {\tilde j}({\bar z}) = ig(z)^*{\tilde T}({\bar z}).
\eea
For any holomorphic function $g(z)$ these are conserved.
These currents lead to the Ward identity
\bea
\delta X^\mu_i(w, {\bar w})=-\epsilon \bigg{(} {\rm Res}_{z\rightarrow w}
g(z)T(z)X^\mu_i(w, {\bar w})+{\bar {\rm Res}}_{{\bar z}\rightarrow {\bar w}}
g(z)^*{\tilde T}({\bar z})X^\mu_i(w, {\bar w})\bigg{)},
\eea
where ``Res'' and ``${\bar {\rm Res}}$'' are coefficients of 
$(z-w)^{-1}$ and $({\bar z}-{\bar w})^{-1}$, respectively.
From the OPEs (31) and (32) and the Ward identity (35) we obtain 
the conformal transformations
\bea
\delta X^\mu (w, {\bar w})= -\epsilon [ g(w) \partial_w X^\mu(w)
+g(w)^* \partial_{\bar w}X^\mu(\bar w)]
\nonumber\\
-\frac{\alpha' \epsilon}
{2(\beta-\lambda^2)}(\beta V^\mu -\lambda U^\mu)
[\partial_w g(w)+\partial_{\bar w} g(w)^*],
\eea
\bea
\delta Y^\mu (w, {\bar w})= -\epsilon[ g(w) \partial_w Y^\mu(w)
+g(w)^* \partial_{\bar w}Y^\mu(\bar w)]
\nonumber\\
-\frac{\alpha' \epsilon}
{2(\beta-\lambda^2)}(-\lambda V^\mu +U^\mu)
[\partial_w g(w)+\partial_{\bar w} g(w)^*].
\eea
Due to the inhomogeneous parts, originated from $V^\mu$, $U^\mu$,
$\beta$ and $\lambda$,
the fields $X^\mu$ and $Y^\mu$ do not transform as conformal tensor.
These parts also indicate that these transformations are not
infinitesimal coordinate transformations $\delta z = \epsilon g(z)$ and
$\delta {\bar z} =\epsilon g(z)^*$.
\section{Mode Expansions}

Now we express some quantities of the model in terms of the oscillators
of $X^\mu$ and $Y^\mu$. The OPEs (31) and (32) give
\bea
&~& T(z)\partial_w X^\mu(w) \sim \frac{1}{(z-w)^3}
\frac{\alpha'}{\beta-\lambda^2}
(\beta V^\mu -\lambda U^\mu) + \frac{1}{(z-w)^2}\partial_w X^\mu(w)
+ \frac{1}{z-w}\partial^2_w X^\mu(w),
\nonumber\\
&~& {\tilde T}({\bar z})\partial_w X^\mu(w) \sim 0 ,
\eea
\bea
&~& T(z)\partial_w Y^\mu(w) \sim \frac{1}{(z-w)^3}
\frac{\alpha'}{\beta-\lambda^2}
(-\lambda V^\mu +U^\mu) + \frac{1}{(z-w)^2}\partial_w Y^\mu(w)
+ \frac{1}{z-w}\partial^2_w Y^\mu(w),
\nonumber\\
&~& {\tilde T}({\bar z})\partial_w Y^\mu(w) \sim 0 .
\eea
Thus, the conformal weights of $\partial_z X^\mu (z)$ 
and $\partial_z Y^\mu (z)$ are
\bea
&~& h_{\partial X} =1\;\;\;,\;\;\; {\tilde h}_{\partial X}=0,
\nonumber\\
&~& h_{\partial Y} =1\;\;\;,\;\;\; {\tilde h}_{\partial Y}=0.
\eea
According to these conformal weights, we obtain the Laurent expansions
\bea
&~& \partial_z X^\mu_i (z)=-i\sqrt{\frac{\alpha'}{2}} \sum^\infty_{m=-\infty}
\frac{\alpha^\mu_{(i)m}}{z^{m+1}},
\nonumber\\
&~& \partial_{\bar z} X^\mu_i({\bar z}) =-i\sqrt{\frac{\alpha'}{2}} 
\sum^\infty_{m=-\infty}\frac{{\tilde \alpha}^\mu_{(i)m}}{{\bar z}^{m+1}}.
\eea
Single-valuedness of $X^\mu$ and $Y^\mu$  implies that 
\bea
\alpha^\mu_{(i)0}={\tilde \alpha}^\mu_{(i)0} =\sqrt{\frac{\alpha'}{2}}
p^\mu_i,
\eea
where $p^\mu_i$ is the linear momentum.
Now integration of the expansions (41) gives the closed string solution
\bea
X^\mu_i(z,{\bar z})=x^\mu_i -i\frac{\alpha'}{2}p^\mu_i \ln|z|^2
+ i\sqrt{\frac{\alpha'}{2}}\sum^\infty_{m\neq 0}\frac{1}{m}
\bigg{(} \frac{\alpha^\mu_{(i)m}}{z^m}+
\frac{{\tilde \alpha}^\mu_{(i)m}}{{\bar z}^m}\bigg{)}.
\eea
Reality of $X^\mu_i$ implies that $\alpha^{\mu\dagger}_{(i)m}
=\alpha^\mu_{(i)(-m)}$ and ${\tilde \alpha}^{\mu\dagger}_{(i)m}
={\tilde \alpha}^\mu_{(i)(-m)}$.

The expansions (41) also lead to
\bea
&~& \alpha^\mu_{(i)m}=\sqrt{\frac{2}{\alpha'}}\oint_C \frac{dz}{2\pi}
z^m \partial_z X^\mu_i(z) ,
\nonumber\\
&~& {\tilde \alpha}^\mu_{(i)m}=-\sqrt{\frac{2}{\alpha'}}
\oint_{\tilde C} \frac{d{\bar z}}{2\pi}
{\bar z}^m \partial_{\bar z} X^\mu_i({\bar z}) ,
\eea
where $C$ in the $z$-plane and ${\tilde C}$ in the ${\bar z}$-plane
are counterclockwise.
Therefore, by using the OPEs $\partial_z X^\mu_i(z)\partial_w X^\nu_j(w)$
and $\partial_{\bar z} X^\mu_i({\bar z})\partial_{\bar w} X^\nu_j({\bar w})$
we obtain
\bea
&~& [\alpha^\mu_{(i)m} , \alpha^\nu_{(j)n}]
=[{\tilde \alpha}^\mu_{(i)m} , {\tilde \alpha}^\nu_{(j)n}]=
\frac{2}{\alpha'}m\eta^{\mu\nu}Q_{ij}\delta_{m,-n},
\nonumber\\
&~& [\alpha^\mu_{(i)m} , {\tilde \alpha}^\nu_{(j)n}]=0,
\nonumber\\
&~& [x^\mu_i , p^\nu_j] =\frac{2i}{\alpha'}\eta^{\mu\nu}Q_{ij}.
\eea
For $\lambda \neq 0$ we observe that the oscillators of $X^\mu$ 
do not commute with the oscillators of $Y^\mu$. 

In terms of the oscillators the nonzero elements of the 
energy-momentum tensor find the forms
\bea
&~& T(z) = -\frac{\alpha'}{2}\eta_{\mu\nu}\Lambda_{ij}
\sum^\infty_{m=-\infty}\sum^\infty_{n=-\infty} \frac{1}{z^{m+n+2}}
:\alpha^\mu_{(i)m}\alpha^\nu_{(j)n}:
+i \sqrt{\frac{\alpha'}{2}}V_\mu^i\sum^\infty_{m=-\infty}
\frac{m+1}{z^{m+2}}\alpha^\mu_{(i)m},
\nonumber\\
&~& {\tilde T}({\bar z}) = -\frac{\alpha'}{2}\eta_{\mu\nu}\Lambda_{ij}
\sum^\infty_{m=-\infty}\sum^\infty_{n=-\infty} \frac{1}{{\bar z}^{m+n+2}}
:{\tilde \alpha}^\mu_{(i)m}{\tilde \alpha}^\nu_{(j)n}:
+i \sqrt{\frac{\alpha'}{2}}V_\mu^i\sum^\infty_{m=-\infty}
\frac{m+1}{{\bar z}^{m+2}}{\tilde \alpha}^\mu_{(i)m}.
\eea
The Virasoro operators are
\bea
&~& L_m = \oint_C \frac{dz}{2\pi i} z^{m+1} T(z),
\nonumber\\
&~& {\tilde L}_m = \oint_{\tilde C} \frac{d{\bar z}}{2\pi i} {\bar z}^{m+1} 
{\tilde T}({\bar z}).
\eea
In terms of the oscillators they take the forms
\bea
&~& L_m = -\frac{\alpha'}{2} \Lambda_{ij}
\eta_{\mu\nu} \sum^\infty_{n=-\infty}
:\alpha^\mu_{(i)m-n}\alpha^\nu_{(j)n}: 
+i\sqrt{\frac{\alpha'}{2}}(m+1)V_\mu^i \alpha^\mu_{(i)m},
\nonumber\\
&~& {\tilde L}_m = -\frac{\alpha'}{2} \Lambda_{ij}\eta_{\mu\nu} 
\sum^\infty_{n=-\infty}
:{\tilde \alpha}^\mu_{(i)m-n}{\tilde \alpha}^\nu_{(j)n}: 
+i\sqrt{\frac{\alpha'}{2}}(m+1)V_\mu^i {\tilde \alpha}^\mu_{(i)m}.
\eea
Using the standard methods one can show that 
the normal ordering constant for all $L_m$ and ${\tilde L}_m$ is zero.
According to the equations (45), or the standard form of the OPEs $TT$
and ${\tilde T}{\tilde T}$, $i.e.$ the equations (29),
the Virasoro algebra also has the standard form
\bea
&~& [L_m , L_n]=(m-n)L_{m+n} +\frac{c}{12}(m^3-m)\delta_{m,-n},
\nonumber\\
&~& [{\tilde L}_m , {\tilde L}_n]=(m-n){\tilde L}_{m+n} 
+\frac{{\tilde c}}{12}(m^3-m)\delta_{m,-n},
\nonumber\\
&~& [L_m , {\tilde L}_n]=0.
\eea

The Hamiltonian of the system is given by
\bea
H=L_0+{\tilde L}_0 -\frac{c+{\tilde c}}{24}.
\eea
Thus, the equations (27) and (48) express this Hamiltonian in terms of
the oscillators and the parameters of the model
\bea
H=&~& \frac{1}{2} \alpha' (p_1.p_1 +2\lambda p_1.p_2 + \beta p_2 . p_2
+ 2i V.p_1+2i U.p_2)
\nonumber\\
&~& 
-\alpha'\Lambda_{ij} \eta_{\mu\nu} \sum_{n=1}^\infty (\alpha^\mu_{(i)(-n)}
\alpha^\nu_{(j)n}+{\tilde \alpha}^\mu_{(i)(-n)}{\tilde \alpha}^\nu_{(j)n})
\nonumber\\
&~& -\frac{\alpha'}{2(\beta-\lambda^2)}(\beta V.V -2\lambda V.U +U.U)
-\frac{D}{6},
\eea
where the symmetry of $\eta_{\mu\nu}$ and $\Lambda_{ij}$ were
introduced.

For the open string there are
\bea
\alpha^\mu_{(i)m}={\tilde \alpha}^\mu_{(i)m}\;\;\;,\;\;\;
\alpha^\mu_{(i)0}={\tilde \alpha}^\mu_{(i)0}= \sqrt{2\alpha'}p^\mu_i,
\eea
and hence the solution is
\bea
X^\mu_i(z,{\bar z})=x^\mu_i -i\alpha'p^\mu_i \ln|z|^2
+ i\sqrt{\frac{\alpha'}{2}}\sum^\infty_{m\neq 0}
\frac{\alpha^\mu_{(i)m}}{m}
\bigg{(}\frac{1}{z^m}+\frac{1}{{\bar z}^m}\bigg{)}.
\eea
The corresponding energy-momentum tensor and Virasoro operators are
given by the first equations of (46) and (48). Thus, the associated 
Virasoro algebra also is described by the first equation of (49).

Note that we imposed the boundary conditions of the closed string and
open string on $Y^\mu (\sigma , \tau)$. However, $Y^\mu (\sigma , \tau)$
may be neither closed nor open. Assuming closeness or openness
for $Y^\mu (\sigma , \tau)$, the worldsheet fields 
$(X^\mu (\sigma , \tau)\;, Y^\mu (\sigma , \tau))$ find four configurations:
(closed , closed), (open , open), (open , closed) and (closed , open).
We considered the first and the second cases. The third and the fourth
cases also can be investigated in the same way.
\section{Conclusions and Summary}

We studied a CFT model with two kinds of the bosonic degrees of freedom
$X^\mu$ and $Y^\mu$, which
interact kinetically with each other. For each kind of these fields we 
introduced a linear dilaton field. 

Using the path integral formalism, we obtained the OPEs $XX$, $XY$ and
$YY$. These OPEs enabled us to introduce a general definition for the 
OPEs. We observed that the $TT$ and ${\tilde T}{\tilde T}$ OPEs of the model
have the standard forms. 
Due to the vectors $V^\mu$ and $U^\mu$, which define the dilatons, the 
OPEs $TX$, $TY$, ${\tilde T}X$ and ${\tilde T}Y$ have deviations from the 
standard forms of them.
The central charge of the model depends on the spacetime
dimension, the parameters of the theory and the vectors $V^\mu$ and
$U^\mu$. A vanishing conformal
anomaly and hence a desirable dimension for the spacetime 
can be achieved by tuning these variables.

Putting away the interacting terms of the 
action, the model split into two copies of the linear dilaton CFT.
The splitting also occurs for the energy-momentum tensor and hence
for the central charge.

Using the conserved currents, associated to the conformal symmetry, 
the conformal transformations of the fields $X^\mu$ and $Y^\mu$ have
been extracted. Therefore, the vectors 
$V^\mu$ and $U^\mu$ and also the parameters
of the model indicate that $X^\mu$ and $Y^\mu$ are not conformal tensors.
That is, these transformations are not pure
coordinate transformations.

According to the mode expansions of $X^\mu$ and $Y^\mu$, we obtained
the oscillator-algebra of the model. Due to the nonzero coupling constant
$\lambda$ the oscillators of $X^\mu$ do not commute with the oscillators 
of $Y^\mu$. We expressed the energy-momentum tensor and the 
Virasoro operators in terms of the oscillators. 
We observed that the Virasoro operators also form the standard algebra.

\end{document}